\documentclass[final]{IEEEtran}

\usepackage{amsthm,amssymb,graphicx,multirow,amsmath,color,amsfonts}
\usepackage[update,prepend]{epstopdf}
\usepackage[noadjust]{cite}
\usepackage[latin1]{inputenc}
\usepackage{tikz}
\usetikzlibrary{arrows,calc}		
\usepackage{bbm} 
\usepackage{pdfpages}
\usepackage{tabulary}
\usepackage{multirow}
\usepackage{comment}
\usepackage{subfigure,array}
\usepackage[linesnumbered,ruled]{algorithm2e}
\usepackage{setspace}	

\def\U{{\Phi_{{\rm u}_0}}}

\def\nb0{{\mathbf{0}}}
\def\nb1{{\mathbf{1}}}

\def\ncalB{{\mathcal{B}}}

\def\nbbE{{\mathbb{E}}}

\def\nbbP{{\mathbb{P}}}

\def\nbbR{{\mathbb{R}}}

\def\nbbZ{{\mathbb{Z}}}

\newtheorem{lemma}{Lemma}

\newtheorem{ndef}{Definition}

\newtheorem{theorem}{Theorem}

\newtheorem{cor}{Corollary}

\newtheorem{remark}{Remark}
\newtheorem{assumption}{Assumption}
	
\def\E{\mathbb{E}}

\def\P{\mathbb{P}}
\def\pc{\mathtt{P_c}}
\def\pr{\mathtt{P_r}}

\def\R{\mathbb{R}}

\def\sinr{\mathtt{SINR}}			

\def\sir{\mathtt{SIR}}

\def\matern{Mat\'ern\ }

\title{Load on the Typical Poisson Voronoi Cell with Clustered User Distribution}
\author{Chiranjib Saha and Harpreet S. Dhillon
\thanks{The authors are with Wireless@VT, Department of ECE, Virgina Tech, Blacksburg, VA, USA. Email: \{csaha,  hdhillon\}@vt.edu. 

The support of the US National Science Foundation (Grant CNS-1617896)  is gratefully acknowledged. 
}}

\begin{document}
\maketitle
\begin{abstract}
In this letter, we characterize the distribution of the number of users associated with the typical base station (BS),  termed the {\em typical cell load}, in a cellular network where the BSs are distributed as a homogeneous Poisson point process (PPP) and the users are distributed as an independent Poisson cluster process (PCP). In this setting, we derive the exact expressions for the first two moments of the typical cell load. Given the computational complexity of evaluating the higher moments, we  derive easy-to-use  approximations for the probability generating function (PGF) of the typical cell load, which can be inverted to obtain the probability mass function  (PMF). 
\end{abstract}
\section{Introduction}
A vast majority of the existing literature on the analysis of cellular networks using stochastic geometry focuses on the distribution of downlink signal-to-interference-and-noise-ratio ($\sinr$) under a variety of settings~\cite{OffloadingSingh,Rate6658810,saha20173gpp,chetlur2019coverage}. While this is important for  evaluating the downlink coverage of the network, the  $\sinr$ distribution by itself is not sufficient to compute the distribution of the  effective downlink rate perceived by the users, which is an equally important metric.  In order to derive the rate distribution of the typical user, we additionally need information about the fraction of resources allocated to that user, which in turn depends upon the load (number of users served) on its serving BS~\cite{OffloadingSingh}. Naturally, load characterization further depends upon the user distribution. While this problem is well-studied for the canonical  PPP-based models (where both user and BS locations are modeled as independent PPPs), the same is not true for the recently developed  PCP-based models for cellular networks~\cite{saha20173gpp}. As a step towards this direction, we characterize load on the typical cell of a PCP-based cellular network model in which the BSs follow a PPP while the users are distributed as an independent PCP. 

{\em Prior Art.} The distributions of the load on the typical cell and the zero cell (i.e. the cell containing the origin) for the cannonical  PPP-based  models are  well-known in the literature~\cite{OffloadingSingh,Rate6658810}. However, the analysis becomes  intractable for the {\em non-PPP} models, i.e. when the PPP assumption on either of the distributions of BSs or users is relaxed. 
 In~\cite{george2018distribution}, the load distribution is characterized assuming PPP-distributed users but a general distribution of the BSs. 
In \cite{chetlur2019coverage}, the authors derive the load distributions assuming that the BSs are distributed as PPP and the users are distributed as a Cox process driven by a Poisson line process. While these works evince the    
 tractability of load distributions for the non-PPP models in general, the analyses  do  not simply extend  to the PCP-based  models developed in~\cite{saha20173gpp}. The current paper presents the first work towards the characterization of load distributions for the PCP-based models.  
 
{\em Contributions.} We consider a cellular network where the BSs are distributed as a  homogeneous PPP and the users are distributed as an independent PCP. For this network, we derive the first two moments (equivalently, the mean and variance) of the typical {\em cell load}, which is defined as the number of points of PCP falling in the  typical cell of the Poisson Voronoi (PV) tessellation generated by the BS PPP.  The key enabling step is the  derivation of the $n^{th}$ moment of typical cell load for a general user point process (PP), whose exact expression is derived in Theorem~\ref{thm::general::moment}. As a special case, we evaluate the first and second moments of cell load  when the user PP is a PCP (Lemma~\ref{lemma::first::two::moments}). To the best of our knowledge, this is the first result on the variance of the cell load for a PCP-based  cellular model.  While these exact results for the moments are key contributions by themselves,  it is unfortunately not very computationally efficient to evaluate these expressions for  $n>2$. For this reason,  we provide an alternate formulation of the load   PGF by approximating the typical cell as a circle with  the same area. We  then obtain  an easy-to-use expression for the  PMF of the typical cell load  by inverting the  PGF. After verifying the accuracy of the analysis with Monte Carlo simulations, we consider the downlink of the cellular network as a case study and   apply this PMF to compute the rate coverage of a randomly chosen user in the typical cell.    

{\em Notations.} (i) We denote a PP and its associated counting measure by the same 
notation, i.e.,  if $\Phi$ denotes a PP, then $\Phi(A)$ denotes the number of points of $\Phi$
falling in $A \in \mathfrak{B}_{\R^2}$, where $\mathfrak{B}_{\R^2}$ denotes the Borel-$\sigma$ algebra in $\R^2$, (ii) $v_2(\cdot)$ denotes the Lebesgue measure  in $\R^2$ (i.e., for a set $B \in \mathfrak{B}_{\R^2}$, $v_2(B)$ denotes the area of $B$),  (iii) $b({\bf x},{\tt R})$ denotes a disc of radius $\tt R$ centered at ${\bf x}\in\R^2$,  (iv) the position vector of a point  in $\R^2$ is denoted as boldface (such as ${\bf x}$), (v) ${\bf 1}(\cdot)$ denotes the indicator function, and (vi) $A_{\rm u}({\tt R}_1,{\tt R}_2,r)$ and $A_{\rm i}({\tt R}_1,{\tt R}_2,r)$ denote the areas of union and intersection of two discs of radii ${\tt R}_1$ and ${\tt R}_2$, whose  centers are separated by a distance $r$. 
\section{System Model}\label{sec::system::model}
We consider a cellular network where the BSs are distributed as a stationary PPP $\Phi_{\rm b}\subset\R^2$ with intensity $\lambda_{\rm b}>0$. The users are assumed to be distributed as another independent PP $\Phi_{\rm u}$. If each user associates with the BS which provides maximum average power,  the association cells of the network form the PV tessellation generated by $\Phi_{\rm b}$~\cite{OffloadingSingh}.  The {\em typical}  association cell centered at ${\bf x}\in \R^2$  is  defined as: ${\cal C}_{\bf x}:=$
 \begin{align}\label{eq::definition::cell}
& \{{\bf y}\in \R^2: \|{\bf y}-{\bf x}\|\leq \|{\bf y}-{\bf t}\|,\forall\ {\bf t}\in\Phi_{\rm b}\}|{\bf x}\in\Phi_{\rm b}.
\end{align}
Note that since $\Phi_{\rm b}$ is a random measure, ${\cal C}_{\bf x}$ is a random closed subset of $\R^2$. Recalling the equivalence of a PP and random counting measure,  $\Phi_{\rm u}({\cal C}_{\bf x})$ is the number of users associated with the BS at ${\bf x}$, equivalently  the  load on the BS at ${\bf x}$. We are interested in characterizing the distribution of the load on the typical BS (termed the typical cell load).  Since $\Phi_{\rm b}$ is stationary (i.e. translation-invariant),  the typical BS can be assumed to be located at the origin (`$o$'). Thus, the typical cell load can be denoted as 
$\U\triangleq \Phi_{\rm u}({\cal C}_{o})$. When $\Phi_{\rm u}$ is a stationary PPP, the  PMF of $\U$ is well-known in the literature~\cite{OffloadingSingh}. However, not much is known if the user distribution is not a PPP. In this letter, we derive the distribution of $\U$ when $\Phi_{\rm u}$ is distributed as a PCP independent of $\Phi_{\rm b}$. In the rest of this section, we will 
introduce PCP and its special cases of interest. 
\begin{ndef}[PCP]\label{def::PCP} A PCP $\Phi_{\rm u}(\lambda_{\rm p},\bar{m},f)$ is defined as 
$\Phi_{\rm u} = \bigcup\limits_{{\bf z}\in \Phi_{\rm p}} {\bf z}+ {\cal B}^{\bf z},$
where $\Phi_{\rm p}$ is the parent PPP with intensity $\lambda_{\rm p}$ and ${\cal B}^{\bf z}$ denotes the offspring  PP centered at ${\bf z}\in\Phi_{\rm p}$. The offspring PP is defined as an independently and identically distributed  (i.i.d.) sequence of   random  vectors $\{{\bf s}\in {\cal B}^{\bf z}\}$ where $\bf s$ follows a probability density function (PDF) $f({\bf s})$ and ${\cal B}^{\bf z}(\nbbR^2)\sim{\rm Poisson}(\bar{m})$.    
\end{ndef}
A PCP is a stationary PP and hence has constant intensity $\lambda_{\rm u}=\bar{m}\lambda_{\rm p}$~\cite[Section~6.4]{haenggi2012stochastic}. We will use the stationarity property of PCP to derive our main results in the next section.  
In this letter, we focus on two well-known special cases of PCP: (i)  the Thomas cluster process (TCP) and (ii) the \matern cluster process (MCP), which are defined as follows. 
\begin{ndef}[TCP]\label{def::TCP}
A PCP  $\Phi_{\rm u}\ (\lambda_{{\rm p}},\bar{m},f)$ is called a TCP if 
 the offspring points in ${\cal B}^{\bf z}$ are distributed normally around ${\bf z}$, i.e., $f({\bf s}) = \frac{1}{2\pi\sigma^2}e^{-\frac{\|{\bf s}\|^2}{2\sigma^2}}$. Here $\sigma^2$ is the cluster variance. 
\end{ndef}
\begin{ndef}[MCP]\label{def::MCP}A PCP  $\Phi_{\rm u}\ (\lambda_{{\rm p}},\bar{m},f)$ is called an MCP if the distribution of the offspring points in $\ncalB^{\bf z}$ is uniform within  $b(o,r_{{\rm d}})$. 
Hence, $
f({\bf s}) =f(s,\theta_s) = \frac{2s}{{\tt R}^2}{\times\frac{1}{2\pi}}, 0\leq s\leq {\tt R},0< \theta_s\leq 2\pi.$
\end{ndef}
 If the offspring points are isotropically distributed around the cluster center,  the joint PDF  $f(s,\theta_s) = \tilde{f}(s)\frac{1}{2\pi}$, where $\tilde{f}(\cdot)$ is the marginal PDF of the radial  coordinate. Then the PDF  of the distance of a point of $\Phi_{\rm u}$ from the origin given its cluster center at ${\bf z}\in \Phi_{\rm p}$ is given by:  
$f_{{\rm d}}(r|{\bf z}) \equiv f_{{\rm d}}(r|\|{\bf z}\|)$.
 We now provide the conditional distance distributions of TCP and MCP. 
  When $\Phi_{\rm u}$ is a TCP, the conditional distance  distribution is Rician with PDF:
 \begin{align}\label{eq::marginal::dist::tcp}
 \scalebox{0.98}{$
f_{{\rm d}}(x|z)
= \frac{x}{ \sigma^2} \exp\left(-\frac{x^2+z^2}{2 \sigma^2}\right) I_0\left(\frac{x z}{\sigma^2}\right),  x,z\geq 0,$}
\end{align} where $I_0(\cdot)$ is the modified Bessel function of the first kind with order zero.
When $\Phi_{\rm u}$ is a MCP, 
$f_{\rm d}(x|z) =$
\begin{align}\label{eq::chi_definition}
&\scalebox{0.95}{$\chi^{(1)}(x,z) = \frac{2 x}{{\tt R}^2}, 0\leq x
\leq {\tt R}-z, 0\leq z\leq {\tt R},$} \\  
&\scalebox{0.95}{$\chi^{(2)}(x,z) = \frac{2 x}{\pi {\tt R}^2}\arccos\bigg(\frac{x^2+z^2-{\tt R}^2}{2xz}\bigg), |{\tt R}-z|<x\leq {\tt R}+z.\notag$}
\end{align}
\begin{figure}
\centering
\subfigure[$\Phi_{\rm u}$ is TCP.]{
\includegraphics[scale=0.45]{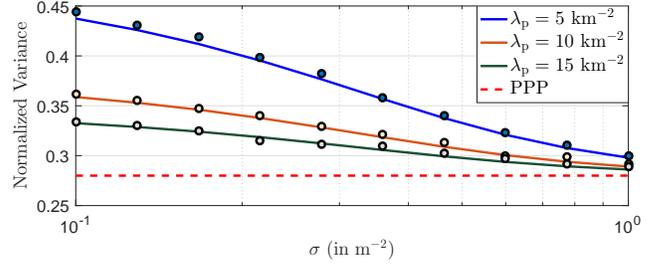}}
\vspace{0.1cm}
\subfigure[$\Phi_{\rm u}$ is MCP.]{
\includegraphics[scale=0.45]{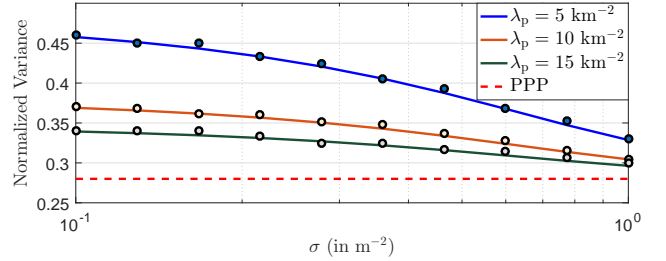}}
\caption{Normalized variance of $\U$ ($\lambda_{\rm b} = 1\ \text{km}^{-2}$). The markers denote the values obtained by Monte Carlo simulation.}\label{fig::variance}
\end{figure}
\begin{figure}
\centering
\subfigure[$\Phi_{\rm u}$ is TCP]{
\includegraphics[scale=0.45]{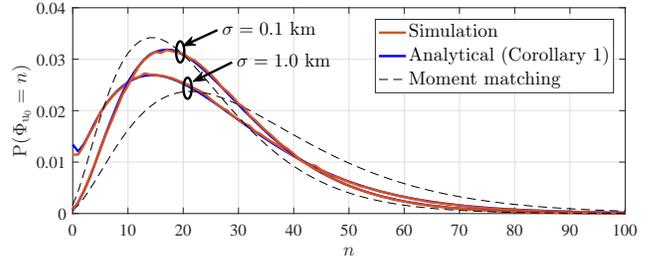}}
\subfigure[$\Phi_{\rm u}$ is MCP]{
\includegraphics[scale=0.45]{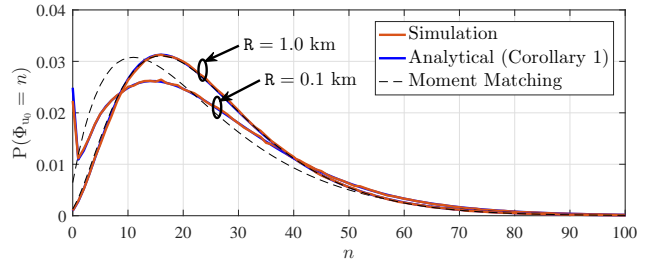}}
\caption{Load on ${\cal C}_o$: comparison of the proposed PMF and the actual PMF when $\Phi_{\rm u}$ is a TCP ($\lambda_{\rm b}=1\ {\rm km}^{-2}$, $\lambda_{\rm p}=5\ {\rm km}^{-2}$, $\bar{m}=5$), $R=1$, $N=128$.}\label{fig::result::typical::cell}
\end{figure}
\section{Moments of the Typical Cell Load}
\label{sec::moments::typical::cell::load}
In this section, we will derive the $n$-th moment of $\U$. We begin with the notion of the moment measure of $\Phi_{\rm u}$. 
\begin{ndef}[Moment measure]\label{def::mu} The $n$-th order moment measure of $\Phi_{\rm u}$ is defined as follows. Given $B_1,\dots, B_n\in\mathfrak{B}_{\R^2}$, 
\begin{multline}\label{eq::mu::def}
\mu^{(n)}(B_1\times\dots B_n):= \E[\Phi_{\rm u}(B_1)\dots \Phi_{\rm u}({B_n})]\\= \E\bigg[\sum\limits_{\substack{{\bf x}_1,\dots,{\bf x}_n\\\in\Phi_{\rm u}}}{\bf 1}({\bf x}_1\in B_1)\dots {\bf 1}({\bf x}_n\in B_n)\bigg].
\end{multline}
 \end{ndef}
  Plugging in $B_i=B,\forall\ i=1,\dots,n$ where $B\in\mathfrak{B}_{\R^2}$, we get $\E[(\Phi_{\rm u}(B))^n]=\mu^{(n)}(B^n)$. However, we cannot simply replace $B$ with ${\cal C}_o$ in order to obtain $\E[(\U)^n]$ since ${\cal C}_o$ is a random closed set. 
 Using the moment measures, the expected summation over the points of the product PP $\Phi_{\rm u}^n\subset\R^{2n}$ can be expressed as an $n$-fold integral over $\R^2$. More formally, for a measurable function $g:\R^{2n}\mapsto\R^+$,
 \begin{multline}\label{eq::Campbell}
 \E\bigg[\sum_{\substack{{\bf x}_1,\dots,{\bf x}_n\\ \in\Phi}}g({\bf x}_1,\dots,{\bf x}_n)\bigg] \\= \int_{\R^{2}}\dots\int_{\R^2}  g({\bf x}_1,\dots,{\bf x}_n)\mu^{(n)}({\rm d}{\bf x}_1,\dots,{\bf x}_n).
 \end{multline}
 \begin{theorem}\label{thm::general::moment}
The $n^{th}$ moment of $\U$ for any general distribution of $\Phi_{\rm u}$ independent of $\Phi_{\rm b}$ can be written as: $\nbbE[({\U})^n] = \E[\mu^{(n)}({\cal C}_o^n)]=$
\begin{equation}\label{eq::exact::moment}
\int\limits_{\R^2}\dots\int\limits_{\R^2}  \exp\bigg(-\lambda_{\rm b}v_2\bigg(\bigcup\limits_{i=1}^{n}b({\bf x}_i,\|{\bf x}_i\|)\bigg)\bigg)\mu^{(n)}({\rm d}{\bf x}_1,\dots,{\rm d}{\bf x
}_n).
\end{equation}
\end{theorem}
\begin{IEEEproof}
Following Definition~\ref{def::mu},  we can write    
\begin{align*}
&\nbbE[({\U})^n]=\E[\mu^{(n)}({\cal C}_o^n)]\\&=\nbbE\bigg[\sum\limits_{\{{\bf x}_i\}\in\Phi_{\rm u}}\prod\limits_{i=1}^{n}{\bf 1}({{\bf x}_i}\in {\cal C}_o)\bigg] \\
&\stackrel{(a)}{=} \nbbE\bigg[\sum\limits_{\{{\bf x}_i\}\in\Phi_{\rm u}}\E_o^!\bigg(\bigcap\limits_{{\bf y}\in\Phi_{\rm b}}\prod\limits_{i=1}^{n}{\bf 1}(\|{\bf x}_i\|<\|{\bf x}_i-{\bf y}\|)\bigg)\bigg]\\
&=\E\bigg[\sum\limits_{\{{\bf x}_i\}\in\Phi_{\rm u}} \nbbP\bigg(\Phi_{\rm b}\bigg(\bigcup\limits_{i=1}^{n}b({\bf x}_i,\|{\bf x}_i\|\bigg) =0\bigg) \bigg]
\\
 &{=} \nbbE\bigg[\sum\limits_{\{{\bf x}_i\}\in\Phi_{\rm u}}\exp\bigg(-\lambda_{\rm b} v_2\bigg(\bigcup\limits_{i=1}^{n}b({\bf x}_i,\|{\bf x}_i\|)\bigg)\bigg)\bigg],
\end{align*}
where $\E_o^!$ in $(a)$ is the expectation with respect to the  reduced Palm distribution of $\Phi_{\rm b}$ which is same as its original distribution, by Slivnyak's theorem~\cite[Theorem~8.3]{haenggi2012stochastic}.  The last step is given by the void probability of PPP~\cite[Section~2.5]{haenggi2012stochastic}. The final expression is obtained by using~\eqref{eq::Campbell}.
\end{IEEEproof}
For stationary PPs, it is possible to simplify \eqref{eq::exact::moment} for $n=1,2$. When $n=1$, $\E[\U] = \lambda_{\rm u}\E[v_2({\cal C}_o)]=\lambda_{\rm u}/\lambda_{\rm b}$, the stationarity of $\Phi_{\rm u}$ and $\Phi_{\rm b}$ imply  $\mu^{(1)}(B)=\lambda_{\rm u}v_2(B)$ and $\E[v_2({\cal C}_o)]=\lambda_{\rm b}^{-1}$, which is the mean area of ${\cal C}_o$~\cite[Theorem~8.3]{haenggi2012stochastic}.  
If $\Phi_{\rm u}$ is a stationary and isotropic PP (which is indeed the case if $\Phi_{\rm u}$ is either TCP or MCP), we can write
\begin{multline*}
\E[\Phi_{\rm u}^2(B)] =  \mu^{(2)}(B^2) = \lambda_{\rm u}v_2(B) \\+ \lambda_{\rm u}
\int_{\R^2}\int_{\R^2} {\bf 1}({\bf x}_1\in B){\bf 1}({\bf x}_2\in B) \varrho^{(2)}(\|{\bf x}_1- {\bf x}_2\|)\: {\rm d}{\bf x}_1 \:{\rm d}{\bf x}_2,
\end{multline*}
where $\varrho^{(2)}(u)$ is called the  second order moment density of $\Phi_{\rm u}$~\cite[Section~6.4]{haenggi2012stochastic}. 
Then  $\E[(\U)^2] = \E[\mu^{(2)}({\cal C}_o^2)] = $
\begin{equation*}
\lambda_{\rm u}\E[v_2({\cal C}_o)]  + \lambda_{\rm u}\int_{\R^2}\int_{\R^2}\exp\big(-\lambda_{\rm b}\times v_2(b({\bf x}_1,\|{\bf x}_1\|)
\end{equation*}
\begin{multline*}
\cup b({\bf x}_2,\|{\bf x}_2\|))\big)\varrho^{(2)}(\|{\bf x}_1-{\bf x}_2\|)\:{\rm d}{\bf x}_1{\rm d}{\bf x}_2
\\= \frac{\lambda_{\rm u}}{\lambda_{\rm b}} +  \lambda_{\rm u}\int_{\R^2}\int_{\R^2}{\cal A}_{\rm u}(\|{\bf x}_1\|,\|{\bf x}_2\|,\|{\bf x}_1-{\bf x}_2\|)\\
\times\varrho^{(2)}(\|{\bf x}_1-{\bf x}_2\|)\:{\rm d}{\bf x}_1\:{\rm d}{\bf x}_2.
\end{multline*} 
Now the  second order moment density of PCP is given by $
\varrho^{(2)}(r) = \lambda_{\rm u}^2(1+\lambda_{\rm p}^{-1}g(r))$ where $g(r) = \int_{\R}f_{\rm d}(r|0)f_{\rm d}(z+r|0)\:{\rm d}z$. This general expression of $\varrho^{(2)}(r)$ can be further simplified when $\U$ is a TCP or a MCP: $\varrho^{(2)}(r)=$
\begin{align}\label{eq::reduced::moment::PCP}
 \begin{cases}
  \lambda_{\rm p}^2\bar{m}^2 + \frac{\lambda_{\rm p}\bar{m}^2}{4\pi\sigma^2}e^{-\frac{r^2}{4\sigma^2}}, &\text{when $\Phi_{\rm u}$ is TCP}\\
  \lambda_{\rm p}^2\bar{m}^2 + {\bf 1}(r\leq 2{\tt R})\frac{\lambda_{\rm p}\bar{m}^2 A_{\rm i}({\tt R},{\tt R},r)}{\pi^2{\tt R}^4},&\text{when $\Phi_{\rm u}$ is MCP}
  \end{cases},
\end{align}
where $A_{\rm i}({\tt R}_1,{\tt R}_2,r) =
$ $ 
{\tt R}_1^2 \arctan\left(\frac{r^2+{\tt R}_1^2-{\tt R}_2^2}{t}\right) + {\tt R}_2^2 \arctan\left(\frac{r^2-{\tt R}_1^2+{\tt R}_2^2}{t}\right) -\frac{t}{2},$ 
with $t = \big(({\tt R_1}+{\tt R}_2+r)({\tt R_1}+{\tt R}_2-r)({\tt R_1}-{\tt R}_2+r)(-{\tt R_1}$ $+{\tt R}_2+r)\big)^{\frac{1}{2}}$ and $0\leq r\leq {\tt R}_1+{\tt R}_2$. Interested readers are advised to refer to  \cite[Section~6.5]{haenggi2012stochastic} for the derivation of these results. 
We now present the mean and variance of $\U$ in the following lemma. 
\begin{lemma}\label{lemma::first::two::moments}
When $\Phi_{\rm u}$ is a PCP, the first two moments of $\U$ are given by:
$
\nbbE[\U] = \frac{\bar{m}\lambda_{\rm p}}{\lambda_{\rm b}},
$
 and $\nbbE[(\U)^2] =$ 
\begin{multline*}
\frac{\lambda_{\rm u}}{\lambda_{\rm b}} +  \int\limits_{0}^{2\pi}\int\limits_{0}^{\infty}\int\limits_{0}^{\infty}\exp(-\lambda_{\rm b}{A}_{\rm u}(x_1,x_2,d(x_1,x_2,\theta))\\\times \varrho^{(2)}(d(x_1,x_2,\theta))x_1x_2\:{\rm d}x_1\:{\rm d}x_2{\rm d}\theta,
\end{multline*}
where $d(x_1,x_2,\theta) := (x_1^2+x_2^2-2x_1x_2\cos\theta)^{\frac{1}{2}}$.
\end{lemma}
\begin{figure*}
${\rm Var}[\U]=$
\begin{align}\label{eq::variance::TCP::MCP}
\begin{cases} 
\frac{\bar{m}^2\lambda_{\rm p}^2}{\lambda_{\rm b}^2}\left( 0.28 + \frac{2}{\lambda_{\rm p}\sigma^2}\int\limits_{0}^{\pi}\int\limits_{0}^{\infty}\int\limits_{0}^{x_1}\exp\left(-A_{\rm u}(x_1,x_2,d(x_1,x_2,\theta))-\frac{d(x_1,x_2,\theta)^2}{4\lambda_{\rm b}\sigma^2}\right)\:x_1\:x_2\:{\rm d}x_2\:{\rm d}x_1 {\rm d}\theta\right),\\
 \qquad\qquad\text{when $\Phi_{\rm u}$ is TCP,}\\
 \frac{\bar{m}^2\lambda_{\rm p}^2}{\lambda_{\rm b}^2}\bigg( 0.28 + \frac{4}{\lambda_{\rm p}{\tt R}^2}\int\limits_{0}^{\infty}\int\limits_{0}^{\pi}\int\limits_{0}^{2{\tt R}}\exp\left(-\lambda_{\rm b} A_{\rm u}(x,(x^2+r^2+2xr\cos\theta),r)\right)A_{\rm i}({\tt R},{\tt R},r)\:x\:r\:{\rm d}r\:{\rm d}\theta\:{\rm d}x\bigg), \\
 \qquad\qquad\text{ when $\Phi_{\rm u}$ is MCP.}
 \end{cases}
 \end{align}
 \hrulefill
\end{figure*}
From Lemma~\ref{lemma::first::two::moments}, we can obtain the  variance of $\U$ which is given by \eqref{eq::variance::TCP::MCP} at the top of the next page. 
We skip the algebraic manipulation due to the lack of space. 
Note that the first term in \eqref{eq::variance::TCP::MCP} denotes the variance of $\U$ if $\Phi_{\rm u}$ is a PPP. Also, $\E[\U]$ is independent of the cluster size (which is $\sigma$ for TCP and $\tt R$ for MCP) and hence is  the  same as the mean cell load under the assumption that $\Phi_{\rm u}$ is a PPP of intensity $\bar{m}\lambda_{\rm p}$.  However,  the variance of $\U$ is higher if $\Phi_{\rm u}$ is a PCP.     The accuracy of \eqref{eq::variance::TCP::MCP} is verified in Fig.~\ref{fig::variance},  where we see that the normalized variance ${\rm Var}[\U]/\E[\U]^2$ for TCP and MCP  matches with the Monte Carlo simulations.  
\begin{remark} \label{rem::other::moments}
Although Theorem~\ref{thm::general::moment} gives the exact expressions of the moments of    $\U$,  we cannot go beyond the  first two moments  since the computation of \eqref{eq::exact::moment} will be limited by the unavailability of the reduced moment measures of PCP for $n\geq 2$ in closed form. This motivates us to formulate a useful approximation  to characterize the distribution of $\U$, which will be presented in the next section.  
\end{remark}
%
With the expressions of the mean and variance of $\U$, 
we now attempt to formulate the PMF of $\U$ using moment matching. To this end, we assume that  $\U$ follows  a negative binomial ($NB$)  distribution, i.e.,  $\U\sim NB(r,t)\Rightarrow \P(\U = n)= {{r+n-1}\choose{n}}(1-t)^r t^n$ (for some $r\in \nbbZ^+, t\in (0,1]$). The intuition behind choosing $NB(r,t)$ is that given any closed subset $B\subset\R^2$,  $\Phi_{\rm u}(B)$   follows a super-Poissonian distribution (i.e. the variance is greater than the mean), and $NB$ is a standard choice for approximating such random variables.  By moment matching, we obtain $\hat{t}=1-\E[\U]/{\rm Var}(\U)$ and $\hat{r} = \lfloor(1-t)\E[\U]/t\rfloor$. In Fig.~\ref{fig::result::typical::cell}, we plot the resulting PMF obtained by moment matching. We observe that for small cluster size (i.e. small $\sigma$ and $\tt R$ for TCP and MCP, respectively), the $NB$ PMF deviates significantly from the empirical PMF of $\U$ obtained from simulation. In particular, the $NB$ distribution significantly underestimates the void probability $\P (\U = 0 )$. Hence the first two moments are not enough to characterize the distribution of $\U$. Since obtaining the exact expressions of  higher order moments is not possible using this route, we provide an alternate formulation for the PMF of $\U$ in the next section.
\section{Derivation of the Load PMF}
\label{sec::load::PMF}
This is the second contribution of the letter, where we start from an approximation of the typical PV cell which eventually leads us to a reasonably accurate characterization of the load PMF. 
  In order to enable the analysis, we approximate the typical cell as a circle with the same area. We formally state this approximation as follows. 
\begin{assumption}\label{assumption::area}
We assume that $\Phi_{\rm u}({\cal C}_o) \approx \Phi_{\rm u}(b(o,R_{\rm c}))$, where $\pi R_{\rm c}^2=v_2({\cal C}_o)$.
\end{assumption}
{While this approximation is inspired by the fact that the large cells in a PV tessellation are circular~ \cite[Theorem~4]{mankar2019distance}, we will demonstrate that this approximation  provides reasonably accurate characterization of the PMF of $\U$ in the non-asymptotic regime as well.   
We first characterize the PGF of $\U$ in the following theorem. 
\begin{theorem}
The PGF of $\U$ is given as: $G_{\U}(\theta)= $
\begin{multline}\label{eq::pmf::load::typical}
 \nbbE\big[\theta^{\U}\big]= \int\limits_{0}^{\infty}\exp\bigg(-2\pi\lambda_{\rm p} \int\limits_{0}^{\infty} \bigg(1-\exp\bigg(-\bar{m}\\
 \times 
 \int\limits_{0}^{r}(1-\theta)f_{\rm d}(u|v){\rm d}{u}\bigg) \bigg)v{\rm d}v\bigg)f_{R_{\rm c}}(r){\rm d}{r},
\end{multline}
where $f_{R_{\rm c}}(r) = {\frac{2\times 3.5^{3.5}}{\Gamma(3.5)}r^{6}\exp\left(-{3.5 r^2}\right)}$. 
\end{theorem}
\begin{IEEEproof}
Following \cite{ferenc2007size}, the random variable $\lambda_{\rm b}v_2({\cal C}_o)$ follows a Gamma distribution with PDF: $f(x;\alpha,\beta) = \frac{\beta^{-\alpha}}{\Gamma(\alpha)}x^{\alpha-1}e^{-\frac{x}{\beta}},\ x>0,$ where $\alpha= 3.5$ and $\beta = 3.5^{-1}$. Since $\pi R_{\rm c}^2 = v_2({\cal C}_o)$, $\sqrt{\pi\lambda_{\rm b}}R_{\rm c}$ follows a Nakagami distribution with PDF $f_{R_{\rm c}}(x;m,\Omega) ={\frac {2m^{m}}{\Gamma (m)\Omega ^{m}}}x^{2m-1}\exp \left(-{\frac {m}{\Omega }}x^{2}\right),\ x>0,$ where $m=3.5$ and $\Omega =1$. 
  We now focus on the conditional PGF of  $\Phi_{\rm u}(b(o,R_{\rm c}))$  given $R_{\rm c}$: $  G_{\Phi_{\rm u}(b(o,R_{\rm c}))}(\theta)=\E\left[\theta^{\Phi_{\rm u}(b(o,R_{\rm c}))}\right]=
  \E\left[\theta^{\sum\limits_{{\bf x}\in \Phi_u}{\bf 1}(\|{\bf x}\|<R_{\rm c})}\right] 
  =\E\left[\prod\limits_{{\bf x}\in \Phi_u}\theta^{{\bf 1}(\|{\bf x}\|<R_{\rm c})}\right].$
  The final step follows from the PGFL of PCP~\cite[Lemma 4]{saha20173gpp} and deconditioning over $R_{\rm c}$. 
\end{IEEEproof}
We now evaluate $G_{\Phi_{\rm u}({\cal C}_o)}(\theta)$ when $\Phi_{\rm u}$ is a TCP (MCP).
\begin{cor}\label{cor::pmf::load}
 When $\Phi_{\rm u}$ is a TCP,  $G_{\Phi_{\rm u}({\cal C}_o)}(\theta)=$
\begin{multline}\label{eq::pmf::load::typical::TCP}
\int\limits_{0}^{\infty}\exp\bigg(-2\pi\lambda_{\rm p} \int\limits_{0}^{\infty} \big(1-\exp\big(-\bar{m}(1-\theta)\\\times\big(1-Q_1({v}{\sigma}^{-1},{r}{\sigma}^{-1})\big)\big)\big)v{\rm d}v\bigg)f_{R_{\rm c}}(r){\rm d}r,
\end{multline}
where $Q_1(\alpha,\beta)= \int_{\beta}^{\infty}ye^{-\frac{y^2+\alpha^2}{2}}I_0(\alpha y){\rm d}{y}$ is the Marcum Q-function. Here $I_0(\cdot)$ is the modified Bessel function of order zero. When $\Phi_{\rm u}$ is an MCP,  $G_{\Phi_{\rm u}({\cal C}_o)}(\theta)$ is given by:
\begin{multline}
\label{eq::pmf::load::typical::MCP}
\int\limits_{0}^{\infty}\exp\bigg(-2\pi\lambda_{\rm p} \int\limits_{0}^{\infty} \big(1-\exp\big(-\bar{m}(1-\theta)\\\times \xi(r,v)\big)\big)v{\rm d}v\bigg)f_{R_{\rm c}}(r){\rm d}{r},
\end{multline}
where 
\begin{multline*}
\xi(r,v) = \frac{1}{{\tt R}^2}\bigg(\big[\min(r,\max({\tt R}-v,0))\big]^2\\
+ \frac{2}{\pi}\int\limits_{\min(r,|{\tt R}-v|)}^{\min(r,{\tt R}+v)} u \arccos\bigg(\frac{u^2+v^2-{\tt R}^2}{2u v}\bigg){\rm d}u \bigg).
\end{multline*}
\end{cor}
Finally the PMF of $\U$, denoted as $\{p_n, n\geq 0\}$, can be obtained by performing the inverse $z$-transform of the PGF which is given by:
\begin{equation}\label{eq::PMF::invert}
p_n = \frac{R^n}{2}\int\limits_{-\pi}^{\pi} G_{\U}(Re^{j\theta})e^{jn\theta}{\rm d}\theta,
\end{equation}
where $R$ is chosen such that $G_{\U}(R e^{j\theta})$ is finite for all $-\pi<\theta<\pi$. For numerical computation, \eqref{eq::PMF::invert} can be  approximated as a summation at $N$ distinct points:
\begin{equation}\label{eq::PMF::approx}
\hat{p}_n = \frac{R^n}{N}\sum\limits_{m=0}^{N-1}G_{\U}(Re^{j2\pi m /N})e^{j2\pi n m/N}.
\end{equation}  
Note that this step is nothing but the inverse discrete Fourier transform (DFT) of $\{G_{\U}(Re^{j2\pi m/N}), m=0,1,\dots,N-1\}$, scaled by $R^n$~\cite{cavers1978fast}. The Matlab scripts for the evaluation of \eqref{eq::PMF::approx} are available in~\cite{SahaILoadcode}. 
In Fig.~\ref{fig::result::typical::cell}, we observe that  $\{\hat{p}_n\}$ closely approximates the true PMF $\P(\U = n)$,  which is  empirically computed from the Monte Carlo simulations  of the network. 
%
\section{Application to  Rate Analysis}\label{sec::application}
In this section, we will apply the PMF of $\U$  to characterize the downlink rate in the cellular network under the system model  defined in Section~\ref{sec::system::model}. In particular, we  evaluate the complementary cumulative density function (CCDF) of rate for a {\em representative user}, which is selected uniformly at random from $\U$ conditioned on the fact that the typical cell has at least one user, i.e., $\U>0$. Assuming that this user is located at ${\bf u}$, the signal-to-interference-ratio ($\sir$) is defined as: 
\begin{equation}\label{eq::sinr::definition}
\sir = \frac{ h_{{o}}\|{\bf u}\|^{-\alpha}}{ \sum\limits_{{\bf x}\in \Phi\setminus\{{o}\}} h_{{\bf x}}\|{\bf x}-{\bf u}\|^{-\alpha}}.
\end{equation}
  Here  $h_{{\bf x}}$ denotes  fading on the link between the representative  user and the BS at ${\bf x}\in\Phi_{\rm b}$, and $\alpha>2$ is the pathloss exponent. We assume Rayleigh fading, i.e., $\{h_{\bf x}\}$ is a sequence of i.i.d.  random variables with $h_{\bf x}\sim \exp(1)$.    Assuming interference-limited network and the system bandwidth (BW) ($W$) is equally partitioned between the  users associated with  a BS, the rate of the representative  user conditioned on $\Phi_{\rm u}({\cal C}_o)>0$ is defined as:   
$
{\tt Rate} = \min\big(\frac{W}{\U}\log(1+\sir),\frac{R_{\rm b}}{\U}\big),$ 
where  $R_{\rm b}$ is the backhaul constraint on the BS  imposed by   the fiber connecting the BS to the network core which can support a maximum rate of $R_{\rm b}\ \text{bps}$. Hence the rate of each user cannot exceed $R_{\rm b}/\U$.  
We define the rate coverage probability as the CCDF of $\tt Rate$: $\pr(\rho) = \P({\tt Rate}>\rho|\U>0)$, where $\rho$ is the target rate threshold. We now provide the expression for the rate coverage in the following theorem. 
\begin{theorem}\label{thm::rate::coverage}
The rate coverage probability for the representative user  is expressed as:
\begin{equation}\label{eq::rate::coverage::probability}
\pr(\rho) \approx \sum\limits_{n=1}^{\lfloor\frac{R_{\rm b}}{\rho}\rfloor}\pc\bigg(2^{\frac{n\rho}{W}-1}\bigg)\frac{\hat{p}_n}{1-\hat{p}_0},
\end{equation}
where $\hat{p}_n$ is obtained from \eqref{eq::PMF::approx} and  $\pc(\tau) = \P(\sir>\tau) $ is the CCDF of $\sir$ that can be expressed as:
\begin{equation}\label{eq::coverage::expression}
\pc(\tau)= \delta^2 \tau^{-\frac{2}{\alpha}}\int\limits_0^{\tau^{\frac{2}{\alpha}}}\frac{\beta(t)^{-2}}{1+t^{\frac{\alpha}{2}}}{\rm d}t,
\end{equation}
where $\beta(t) = t\int\limits_{t^{-1}}^{\infty} \frac{1}{1+u^{\frac{2}{\alpha}}}{\rm d}u$, with $\delta = \frac{9}{7}$. 
\end{theorem}
\begin{IEEEproof}
Given the backhaul constraint, the maximum users that can be supported with a rate $\rho$ is given by $\lfloor R_{\rm b}/\rho\rfloor$. 
 First we note that $\tt Rate$ is a function of  $\sir$ and $\U$, which are in general correlated. However,    the joint distribution of $\sir$ and $\U$ is intractable.  For tractability, we assume that these two random variables are  independent. This is a well-accepted assumption in the literature  that preserves the accuracy of the analysis~\cite[Section~3]{Rate6658810}. Under this assumption, the rate coverage can be expressed as:
 $\pr (\rho)=\P\big(\frac{W}{\Phi_{\rm u}({\cal C}_o)}\log(1+\sir) >\rho|\Phi_{\rm u}({\cal C}_o)>0\big)=$ $\P\bigg(\sir>2^{\frac{\Phi_{\rm u}({\cal C}_o)\rho}{W}}-1\big|\Phi_{\rm u}({\cal C}_o)>0\bigg)=$
\begin{align*}
&= \sum\limits_{n = 1}^{\lfloor \frac{R_{\rm b}}{\rho}\rfloor}\overbrace{\pc(2^{n\rho/W}-1)}^{\text{$\sir$ distribution}}\times \underbrace{\P({\cal C}_o(\Phi_{\rm u})=n|{\cal C}_o(\Phi_{\rm u})>0)}_{\text{load distribution}}.
\end{align*}
The load distribution can be simplified as: $\frac{\P({\cal C}_o(\Phi_{\rm u})=n,{\cal C}_o(\Phi_{\rm u})>0)}{\P ({\cal C}_o(\Phi_{\rm u})>0)}$. Hence we are left with the characterization of  $\pc$, or the CCDF of  $\sir$. Since $\Phi_{\rm u}$ and $\Phi_{\rm b}$ are independent and $\Phi_{\rm u}$ is a stationary distribution {(i.e. the distribution of $\Phi_{\rm u}$ is invariant under translation of its points)}, the representative user is equivalent to a randomly selected point in ${\cal C}_o$. The $\sir$ distribution of this point has been recently  characterized in~\cite{mankar2019downlink}.  The expression of $\pc(\tau)$ in \eqref{eq::coverage::expression} is obtained from \cite[Theorem~2]{mankar2019downlink}. 
\end{IEEEproof}
\begin{figure}
\centering
\subfigure[\label{fig::validation::rate}]{
\includegraphics[scale=0.45]{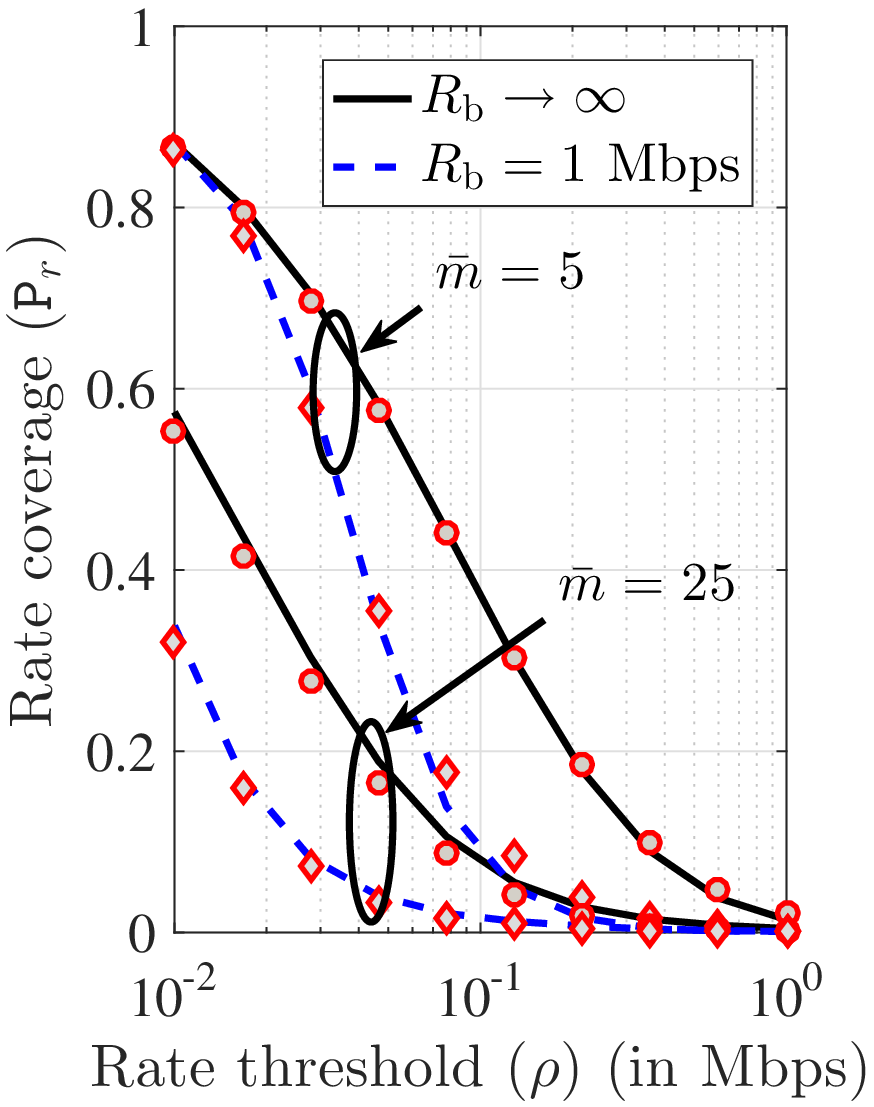}}
\subfigure[\label{fig::rate::cov::vary::sigma}]{
\includegraphics[scale=0.45]{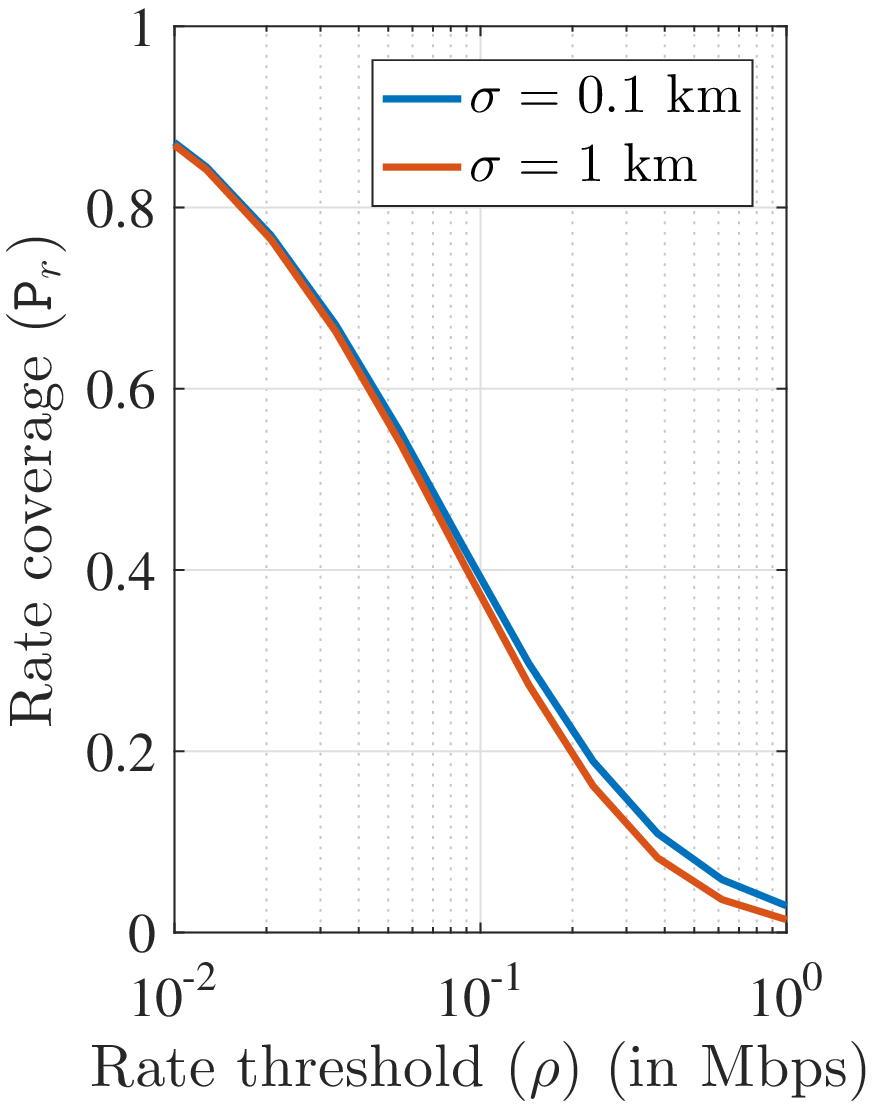}}
\caption{Rate coverage probability: \subref{fig::validation::rate} for different $\bar{m}$ (markers indicate the values obtained from Monte Carlo simulation) and  \subref{fig::rate::cov::vary::sigma} for different $\sigma$ with $R_{\rm b}\to\infty$ ($(\lambda_{\rm b},\lambda_{\rm p})= (1,5) \text{km}^{-2}$ and $W=1$ MHz).}\label{fig::rate::cov}
\end{figure}
We verify the accuracy of Theorem~\ref{thm::rate::coverage} in Fig.~\ref{fig::validation::rate} which exhibits a close match between the analytical and empirical results. Because of the space constraint, we only present the results when $\Phi_{\rm u}$ is a TCP.  We observe that $\pr$ decreases as (i) $\bar{m}$ increases as more number of users share the resources and (ii) $R_{\rm b}$ decreases as it imposes an upper bound on the per user rate. In Fig.~\ref{fig::rate::cov::vary::sigma}, we plot $\pr$ for different $\sigma$ which is a measure of the cluster size. We further observe that $\pr$ is almost invariant to $\sigma$. The reason is that the rate coverage is mostly dominated by the first moment of load (see~\cite[Corollary~1]{OffloadingSingh}) which is independent of the cluster size.

\section{Conclusion}\label{sec::conclusion}
Due to the limitation of PPP in modeling  spatial coupling between the nodes,  there has been  increasing interests in developing  non-PPP models of cellular networks, such as the PCP-based models which capture coupling between the users (such as in hotspots) and between users and BSs~\cite{saha20173gpp}. While the $\sinr$ distribution for the PCP-based  models  is by now well-understood,   the load distribution in these networks has not received much attention. In this letter, we made the first attempt towards this direction by characterizing the distribution of the typical cell load where the BSs are distributed as a homogeneous PPP and the users are distributed as an independent PCP. We also demonstrated the utility of this result by using it to characterize the user rate for a representative user in the typical cell.    

\bibliographystyle{IEEEtran}
\bibliography{Letter_draft_R1.bbl}

\begin{thebibliography}{10}
\providecommand{\url}[1]{#1}
\csname url@samestyle\endcsname
\providecommand{\newblock}{\relax}
\providecommand{\bibinfo}[2]{#2}
\providecommand{\BIBentrySTDinterwordspacing}{\spaceskip=0pt\relax}
\providecommand{\BIBentryALTinterwordstretchfactor}{4}
\providecommand{\BIBentryALTinterwordspacing}{\spaceskip=\fontdimen2\font plus
\BIBentryALTinterwordstretchfactor\fontdimen3\font minus
  \fontdimen4\font\relax}
\providecommand{\BIBforeignlanguage}[2]{{%
\expandafter\ifx\csname l@#1\endcsname\relax
\typeout{** WARNING: IEEEtran.bst: No hyphenation pattern has been}%
\typeout{** loaded for the language `#1'. Using the pattern for}%
\typeout{** the default language instead.}%
\else
\language=\csname l@#1\endcsname
\fi
#2}}
\providecommand{\BIBdecl}{\relax}
\BIBdecl

\bibitem{OffloadingSingh}
S.~Singh, H.~S. Dhillon, and J.~G. Andrews, ``Offloading in heterogeneous
  networks: Modeling, analysis, and design insights,'' \emph{IEEE Trans. on
  Wireless Commun.}, vol.~12, no.~5, pp. 2484--2497, May. 2013.

\bibitem{Rate6658810}
H.~S. Dhillon and J.~G. Andrews, ``Downlink rate distribution in heterogeneous
  cellular networks under generalized cell selection,'' \emph{IEEE Wireless
  Commun. Letters}, vol.~3, no.~1, pp. 42--45, Feb. 2014.

\bibitem{saha20173gpp}
C.~Saha, M.~Afshang, and H.~S. Dhillon, ``3{GPP}-inspired {H}et{N}et model
  using {P}oisson cluster process: Sum-product functionals and downlink
  coverage,'' \emph{IEEE Trans. on Commun.}, vol.~66, no.~5, pp. 2219--2234,
  May 2018.

\bibitem{chetlur2019coverage}
V.~V. Chetlur and H.~S. Dhillon, ``Coverage and rate analysis of downlink
  cellular vehicle-to-everything ({C-V2X}) communication,'' \emph{IEEE Trans.
  on Wireless Commun.}, vol.~19, no.~3, pp. 1738--1753, Mar. 2020.

\bibitem{george2018distribution}
G.~George, A.~Lozano, and M.~Haenggi, ``Distribution of the number of users per
  base station in cellular networks,'' \emph{IEEE Wireless Commun. Letters},
  vol.~8, no.~2, pp. 520--523, 2018.

\bibitem{haenggi2012stochastic}
M.~Haenggi, \emph{Stochastic Geometry for Wireless Networks}.\hskip 1em plus
  0.5em minus 0.4em\relax Cambridge University Press, 2012.

\bibitem{mankar2019distance}
P.~D. Mankar, P.~Parida, H.~S. Dhillon, and M.~Haenggi, ``Distance from the
  nucleus to a uniformly random point in the $0$-cell and the typical cell of
  the {P}oisson-{V}oronoi tessellation,'' 2019, available online:
  arXiv/abs/1907.03635.

\bibitem{ferenc2007size}
J.-S. Ferenc and Z.~N{\'e}da, ``On the size distribution of {P}oisson {V}oronoi
  cells,'' \emph{Physica A: Statistical Mechanics and its Applications}, vol.
  385, no.~2, pp. 518--526, 2007.

\bibitem{cavers1978fast}
J.~Cavers, ``On the fast fourier transform inversion of probability generating
  functions,'' \emph{IMA Journal of Applied Mathematics}, vol.~22, no.~3, pp.
  275--282, 1978.

\bibitem{SahaILoadcode}
C.~Saha and H.~S. Dhillon, ``Matlab code for the computation of the {PMF} of
  the number of points of a {PCP} in a typical cell of a stationary {PPP},''
  available at: {https://github.com/stochastic-geometry/LoadDistributionPCP}.

\bibitem{mankar2019downlink}
P.~D. Mankar, P.~Parida, H.~S. Dhillon, and M.~Haenggi, ``Downlink analysis for
  the typical cell in {P}oisson cellular networks,'' \emph{IEEE Wireless
  Commun. Letters}, vol.~9, no.~3, pp. 336--339, Mar. 2020.

\end{thebibliography}
\end{document}